\documentclass[runningheads]{llncs}
\usepackage{graphicx}
\usepackage{amsmath}
\usepackage{amssymb}

\begin{document}

\title{Uncertain Case Identifiers in Process Mining:\\A User Study of the Event-Case Correlation Problem on Click Data\thanks{We thank the Alexander von Humboldt (AvH) Stiftung for supporting our research interactions.
}}

\author{Marco Pegoraro\orcidID{0000-0002-8997-7517} \and
	Merih Seran Uysal\orcidID{0000-0003-1115-6601} \and
	Tom-Hendrik H\"ulsmann\orcidID{0000-0001-8389-5521} \and
	Wil M.P. van der Aalst\orcidID{0000-0002-0955-6940}}

\authorrunning{M. Pegoraro et al.}
\titlerunning{Event-Case Correlation on Click Data: a User Study}

\institute{Department of Computer Science, RWTH Aachen, Aachen, Germany
	\email{\{pegoraro,uysal,wvdaalst\}@pads.rwth-aachen.de}\\
	\email{tom.huelsmann@rwth-aachen.de}\\
	\url{http://www.pads.rwth-aachen.de/}}

\maketitle

\begin{abstract}
Among the many sources of event data available today, a prominent one is user interaction data. User activity may be recorded during the use of an application or website, resulting in a type of user interaction data often called click data. An obstacle to the analysis of click data using process mining is the lack of a case identifier in the data. In this paper, we show a case and user study for event-case correlation on click data, in the context of user interaction events from a mobility sharing company. To reconstruct the case notion of the process, we apply a novel method to aggregate user interaction data in separate user sessions---interpreted as cases---based on neural networks. To validate our findings, we qualitatively discuss the impact of process mining analyses on the resulting well-formed event log through interviews with process experts.

\end{abstract}

\section{Introduction}\label{sec:introduction}
In the last decades, the dramatic rise of both performance and portability of computing devices has enabled developers to design software with an ever-increasing level of sophistication. Such escalation in functionalities caused a subsequent increase in the complexity of software, making it harder to access for users. The shift from large screens of desktop computers to small displays of smartphones, tablets, and other handheld devices has strongly contributed to this increase in the intricacy of software interfaces. User interface (UI) design and user experience (UX) design aim to address the challenge of managing complexity, to enable users to interact easily and effectively with the software.

In designing and improving user interfaces, important sources of guidance are the records of user interaction data. Many websites and apps track the actions of users, such as pageviews, clicks, and searches. Such type of information is often called \emph{click data}, of which an example is given in Table~\ref{table:data}. These can then be analyzed to identify parts of the interface which need to be simplified, through, e.g., pattern mining, or performance measures such as time spent performing a certain action or visualizing a certain page.

\begin{table}[]
	\scriptsize
	\centering
	\caption{A sample of click data from the user interactions with the smartphone app of a German mobility sharing company. This dataset is the basis for the qualitative evaluation of the method presented in this paper.}
	\begin{tabular}{|c|c|c|c|c|}
		\hline
		\textbf{timestamp}               & \textbf{screen}       & \textbf{user}  & \textbf{team}  & \textbf{os}      \\ \hline
		2021-01-25 23:00:00.939 & \texttt{pre\_booking} & b0b00 & 2070b & iOS     \\ \hline
		2021-01-25 23:00:03.435 & \texttt{tariffs}      & b0b00 & 2070b & iOS     \\ \hline
		2021-01-25 23:00:04.683 & \texttt{menu}         & 3fc0c & 02d1f & Android \\ \hline
		2021-01-25 23:00:05.507 & \texttt{my\_bookings} & 3fc0c & 02d1f & Android \\ \hline
		$\vdots$                & $\vdots$              & $\vdots$      & $\vdots$     & $\vdots$       \\ \hline
	\end{tabular}
	\label{table:data}
	\normalsize
\end{table}

In the context of novel click data analysis techniques, a particularly promising subfield of data science is \emph{process mining}. Process mining is a discipline that aims to analyze event data generated by process executions, to e.g. obtain a model of the process, measure its conformance with normative behavior, or analyze the performance of process instances with respect to time.

Towards the analysis of click data with process mining, a foundational challenge remains: the association of event data (here, user interactions) with a \emph{process case identifier}. While each interaction logged in a database is associated with a user identifier, which is read from the current active session in the software, there is a lack of an attribute to isolate events corresponding to one single utilization of the software from beginning to end. Aggregating user interactions into cases is of crucial importance, since the case identifier---together with the \emph{activity label} and the \emph{timestamp}---is a fundamental attribute to reconstruct a process instance as a sequence of activities (\emph{trace}), also known as \emph{control-flow perspective} of a process instance. A vast majority of the process mining techniques available require the control-flow perspective of a process to be known.

In this paper, we propose a novel case attribution approach for click data. Our method allows us to effectively segment the sequence of interactions from a user into separate cases on the basis of normative behavior. We then verify the effectiveness of our method by applying it to a real-life use case scenario related to a mobility sharing smartphone app. Then, we perform common process mining analyses such as process discovery on the resulting segmented log, and we conduct a user study among business owners by presenting the result of such analyses to process experts from the company. Through interviews with such experts, we assess the impact of process mining analysis techniques enabled by our event-case correlation method.

The remainder of the paper is organized as follows. Section~\ref{sec:related} discusses existing event-case correlation methods and other related work. Section~\ref{sec:method} illustrates a novel event-case correlation method. Section~\ref{sec:qual} describes the results of our method on a real-life use case scenario related to a mobility sharing app, together with a discussion of interviews of process experts from the company about the impact of process mining techniques enabled by our method. Finally, Section~\ref{sec:conclusion} concludes the paper.

\section{Related Work}\label{sec:related}
The problem of assigning a case identifier to events in a log is a long-standing challenge in the process mining community~\cite{DBLP:conf/bpm/FerreiraG09}, and is known by multiple names in literature, including \emph{event-case correlation} problem~\cite{DBLP:conf/er/BayomieCRM19} and \emph{case notion discovery} problem~\cite{DBLP:journals/kais/MurillasRA20}. Event logs where events are missing the case identifier attribute are usually referred to as \emph{unlabeled event logs}~\cite{DBLP:conf/bpm/FerreiraG09}. Several of the attempts to solve this problem, such as an early one by Ferreira et al. based on first order Markov models~\cite{DBLP:conf/bpm/FerreiraG09} or the \emph{Correlation Miner} by Pourmiza et al., based on quadratic programming~\cite{DBLP:journals/ijcis/PourmirzaDG17} are very limited in the presence of loops in the process. Other approaches, such as the one by Bayomie et al.~\cite{DBLP:conf/caise/BayomieAE16} can indeed work in the presence of loops, by relying on heuristics based on activities duration which lead to a set of candidate segmented logs. This comes at the cost of a slow computing time. An improvement of the aforementioned method~\cite{DBLP:conf/er/BayomieCRM19} employs simulated annealing to select an optimal case notion; while still very computationally heavy, this method delivers high-quality case attribution results.

The problem of event-case correlation can be positioned in the broader context of \emph{uncertain event data}~\cite{DBLP:journals/is/PegoraroUA21,DBLP:conf/apn/PegoraroUA21}. This research direction aims to analyze event data with imprecise attributes, where single traces might correspond to an array of possible real-life scenarios. Akin to the method proposed in this paper, some techniques allow to obtain probability distributions over such scenarios~\cite{DBLP:conf/icpm/0001BUA21}.

A notable and rapidly-growing field where the problem of event-case correlation is crucial is \emph{Robotic Process Automation} (RPA), the automation of process activities through software bots. Similar to many approaches related to the problem at large, existing approaches to event-case correlation in the RPA field often heavily rely on unique start and end events in order to segment the log, either explicitly or implicitly~\cite{DBLP:conf/gi/LinnZW18,DBLP:conf/caise/RamirezR0V19,DBLP:conf/icpm/LenoADRMP20}.

The problem of event-case attribution is different when considered on click data---particularly from mobile apps. Normally, the goal is to learn a function that receives an event as an independent variable and produces a case identifier as an output. In the scenario studied in this paper, however, the user is tracked by the open session in the app during the interaction, and recorded events with different user identifier cannot belong to the same process case. The goal is then to subdivide the sequence of interactions from one user into one or more sessions (cases). Marrella et al.~\cite{DBLP:conf/ACMdis/MarrellaC18} examined the challenge of obtaining case identifiers for unsegmented user interaction logs in the context of learnability of software systems, by segmenting event sequences with a predefined set of start and end activities as normative information. They find that this approach cannot discover all types of cases, which limits its flexibility and applicability. Jlailaty et al.~\cite{DBLP:conf/IEEEscc/JlailatyGB17} encounter the segmentation problem in the context of email logs. They segment cases by designing an ad-hoc metric that combines event attributes such as timestamp, sender, and receiver. Their results however show that this method is eluded by edge cases. Other prominent sources of sequential event data without case attribution are IoT sensors: Janssen et al.~\cite{DBLP:conf/icpm/JanssenMKZ20} address the problem of obtaining process cases from sequential sensor event data by splitting the long traces according to an application-dependent fixed length, to find the optimal sub-trace length such that, after splitting, each case contains only a single activity. One major limitation of this approach that the authors mention is the use of only a single constant length for all of the different activities, which may have varying lengths. More recently, Burattin et al.~\cite{DBLP:journals/dke/BurattinKNW19} tackled a segmentation problem for user interactions with a modeling software; in their approach, the segmentation is obtained exploiting eye tracking data.

The goal of the study reported in this paper is to present a method able to rapidly and efficiently segment a user interaction log in a setting where no sample of ground truth cases are available, and the only normative information at disposal is in the form of a link graph relatively easy to extract from a UI. Section~\ref{sec:method} shows the segmentation technique we propose.

\section{Method}\label{sec:method}
In this section, we illustrate our proposed method for event-case correlation on click data. As mentioned earlier, the goal is to segment the sequence of events corresponding to the interactions of every user in the database into complete process executions (cases). In fact, the click data we consider in this study have a property that we need to account for while designing our method: all events belonging to one case are contiguous in time. Thus, our goal is to determine split points for different cases in a sequence of interactions related to the same user. More concretely, if a user of the app produces the sequence of events $\langle e_1, e_2, e_3, e_4, e_5, e_6, e_7, e_8, e_9 \rangle$, our goal is to section such sequence in contiguous subsequences that represent a complete interaction---for instance, $\langle e_1, e_2, e_3, e_4 \rangle$, $\langle e_5, e_6 \rangle$, and $\langle e_7, e_8, e_9 \rangle$. We refer to this as the \emph{log segmentation} problem, which can be considered a special case of the event-case correlation problem. In this context, ``\emph{unsegmented} log'' is synonym with ``unlabeled log''.

Rather than being based on a collection of known complete process instances as training set, the creation of our segmentation model is based on behavior described by a model of the system. A type of model particularly suited to the problem of segmentation of user interaction data---and especially click data---is the \emph{link graph}. In fact, since the activities in our process correspond to screens in the app, a graph of the links in the app is relatively easy to obtain, since it can be constructed in an automatic way by following the links between views in the software. This link graph will be the basis for our training data generation procedure.

We will use as running example the link graph of Figure~\ref{fig:dfg}. The resulting normative traces will then be used to train a neural network model based on the word2vec architecture~\cite{DBLP:conf/nips/MikolovSCCD13}, which will be able to split contiguous user interaction sequences into cases.

\subsection{Training Log Generation}\label{sec:gen}
To generate the training data, we will begin by exploiting the fact that each process case will only contain events associated with one and only one user. Let $L$ be our unsegmented log and $u \in U$ be a user in $L$; then, we indicate with $L_u$ the sub-log of $L$ where all events are associated with the user $u$.

Our training data will be generated by simulating a transition system annotated with probabilities. The construction of a transition system based on event data is a well-known procedure in process mining~\cite{DBLP:journals/sosym/AalstRVDKG10}, which requires to choose an event representation abstraction and a window size (or horizon), which are process-specific. In the context of this section, we will show our method using a sequence abstraction with window size 2. Initially, for each user $u \in U$ we create a transition system $TS_u = (S_u, E_u, T_u, i)$ based on the sequence of user interactions in the sub-log $L_u$. $S_u^{\text{end}} \in S_u$ denotes the final states of $TS_u$. All such transition systems $TS_u$ share the same initial state $i$. To identify the end of sequences, we add a special symbol to the states $f \in S'$ to which we connect any state $s \in S$ if it appears at the end of a user interaction sequence. To traverse the transitions to the final state $f$ we utilize as placeholder the empty label $\tau$.

\begin{figure}
	\centering
	\begin{minipage}[t]{0.48\textwidth}
		\centering
		\includegraphics[width=.5\textwidth]{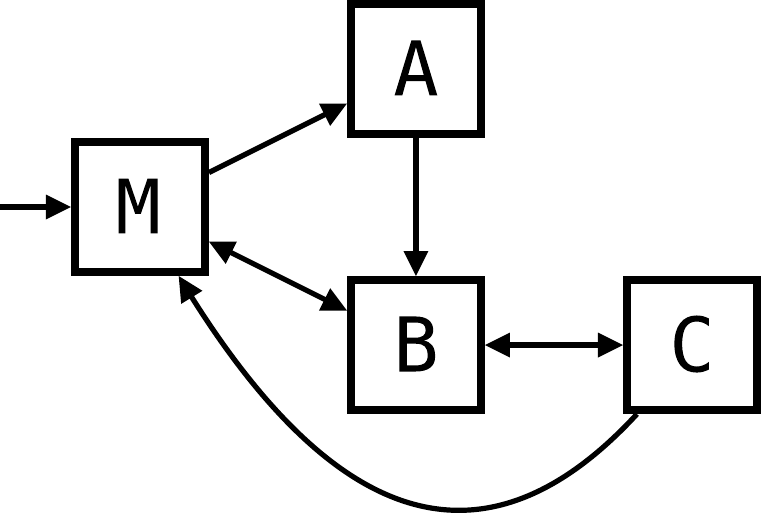}
		\caption{The link graph of a simple, fictional system that we are going to use as running example. From this process, we aim to segment the three unsegmented user interactions $\langle M, A, M, B, C \rangle$, $\langle M, B, C, M \rangle$, and $\langle M, A, B, C \rangle$.}
		\label{fig:dfg}
	\end{minipage}\hfill
	\begin{minipage}[t]{0.48\textwidth}
		\centering
		\includegraphics[width=\textwidth]{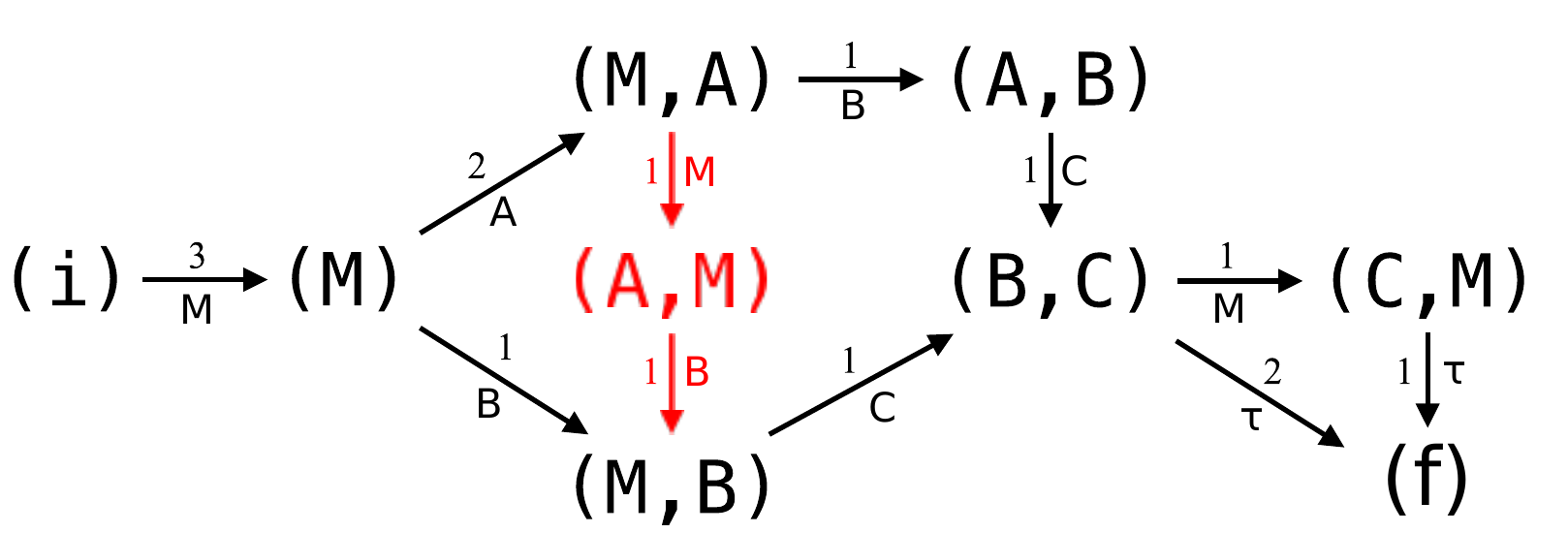}
		\caption{The transition system $TS'$ obtained by the user interaction data of the example (Figure~\ref{fig:dfg}). During the reduction phase, the transition $(M, A)$ to $(A, M)$ is removed, since it is not supported by the link graph ($M$ does not follow $A$). The state $(A, M)$ is not reachable and is removed entirely (in red). Consequently, the reduced transition system $TS$ is obtained.}
		\label{fig:ts}
	\end{minipage}
\end{figure}

We then obtain a transition system $TS' = (S', A, T', i)$ corresponding to the entire log $L$, where $A$ is the set of activity labels appearing in $L$, $S' = \bigcup_{u \in U} S_u$, and $T' = \bigcup_{u \in U} T_u$. Moreover, $S'^{\text{end}} = \bigcup_{u \in U} S_u^{\text{end}}$. We also collect information about the frequency of each transition in the log: we define a weighting function $\omega$ for the transitions $t \in T$ where $\omega(t) = \textit{\# of occurrences of t in L}$. If $t \notin T$, $\omega(t) = 0$. Through $\omega$, it is optionally possible to filter out rare behavior by deleting transitions with $\omega(t) < \epsilon$, for a small threshold $\epsilon$. Figure~\ref{fig:ts} shows a transition system with the chosen abstraction and window size, annotated with both frequencies and transition labels, for the user interactions $L_{u_1} = \langle M, A, M, B, C \rangle$, $L_{u_2} = \langle M, B, C, M \rangle$, and $L_{u_3} = \langle M, A, B, C \rangle$.

In contrast to transition systems that are created based on logs that are segmented, the obtained transition system might contain states that are not reachable and transitions that are not possible according to the real process. Normally, the transition system abstraction is applied on a case-by-case basis. In our case, however, we applied the abstraction to the whole sequence of interactions that is associated with a specific user, consecutive interactions that belong to different cases will be included as undesired transitions in the transition system. In order to prune undesired transitions from the transition system, we exploit the link graph of the system: a transition in the transition system is only valid if it appears in the link graph. Unreachable states are also pruned.

We will assume a sequence abstraction in $TS$. Given a link graph $G = (V, E)$, we define the reduced transition system $TS = (S, A, T, i)$, where $T = \{(\langle \dots, a_1 \rangle, a_2, \langle \dots, a_1, a_2 \rangle) \in T' \mid (a_1, a_2) \in E\}$ and $S = \bigcup_{(s_1, a, s_2) \in t} \{s_1, s_2\}$. Figure~\ref{fig:dfg} shows a link graph for our running example, and Figure~\ref{fig:ts} shows how this is used to reduce $TS'$ into $TS$.

Next, we define probabilities for transitions and states based on the values for $\omega(t)$. Let $T_{\text{out}} \colon S \to \mathcal{P}(T)$ be  $T_{\text{out}}(s) = \{(s_1, a, s_2) \in T \mid s_1 = s\}$; this function returns all outgoing transitions from a given state. The likelihood of a transition $(s_1, a, s_2) \in T$ is then computed with $l_{\text{trans}} \colon T \to [0, 1]$:

$$
l_{\text{trans}}(s_1, a, s_2) = \frac{\omega(s_1, a, s_2)}{\sum\limits_{t_* \in T_{\text{out}}(s_1)} \omega(t_*)}
$$

Note that if $s_1$ has no outgoing transition and $T_{\text{out}}(s_1) = \varnothing$, by definition $l_{\text{trans}}(s_1, a, s_2) = 0$ for any $a \in A$ and $s_2 \in S$. We will need two more supporting functions. We define $l_{\text{start}} \colon S \to [0, 1]$ and $l_{\text{end}} \colon S \to [0, 1]$ as the probabilities that a state $s \in S$ is, respectively, the initial and final state of a sequence:\\

\noindent\begin{minipage}{.5\linewidth}
	$$
	l_{\text{start}}(s) = \frac{\sum\limits_{a \in A}\omega(i, a, s)}{\sum\limits_{\substack{s_* \in S \\ a \in A}} \omega(s_*, a, s)}
	$$
\end{minipage}
\begin{minipage}{.5\linewidth}
	$$
	l_{\text{end}}(s) = \frac{\omega(s, \tau, f)}{\sum\limits_{\substack{s_* \in S \\ a \in A}} \omega(s, a, s_*)}
	$$
\end{minipage}\\

In our running example of Figure~\ref{fig:ts}, $l_{\text{start}}((M)) = \frac{3}{3} = 1$, and $l_{\text{end}}((C, M)) = \frac{1}{3}$. Given a path of states $\langle s_1, s_2, \dots, s_n \rangle$ transitioning through the sequence $\langle (i, a_1, s_1),\allowbreak(s_1, a_2, s_2), \dots, (s_{n - 1}, a_n, s_n), (s_n, \tau, f) \rangle$, we now have the means to compute its probability with the function $l \colon S^* \to [0, 1]$:

$$
l(\langle s_1, s_2, \dots, s_n \rangle) = l_{\text{start}}(s_1) \cdot \prod\limits_{\substack{i = 2}}^n l_{\text{trans}}(s_{i - 1}, a_i, s_{i}) \cdot l_{\text{end}}(s_n)
$$

This enables us to obtain an arbitrary number of well-formed process cases as sequences of activities $\langle a_1, a_2, \dots, a_n \rangle$, utilizing a Monte Carlo procedure. We can sample a random starting state for the case, through the probability distribution given by $l_{\text{start}}$; then, we compose a path with the probabilities provided by $l_{\text{trans}}$ and $l_{\text{end}}$. The traces sampled in this way will reflect the available user interaction data in terms of initial and final activities, and internal structure, although the procedure still allows for generalization. Such generalization is, however, controlled thanks to the pruning provided by the link graph of the system. We will refer to the set of generated traces as the training log $L_T$.

\subsection{Model Training}\label{sec:training}
The training log $L_T$ obtained in Section~\ref{sec:gen} is now used in order to train the segmentation models. The core component of the proposed method consists one or more word2vec models to detect the boundaries between cases in the input log. When applied for natural language processing, the input of a word2vec model is a corpus of sentences which consist of words. Instead of sentences built as sequences of words, we consider traces $\langle a_1, a_2, \dots, a_n \rangle$ as sequences of activities.

\begin{figure}
	\centering
	\begin{minipage}[t]{0.45\textwidth}
		\centering
		\includegraphics[width=\textwidth]{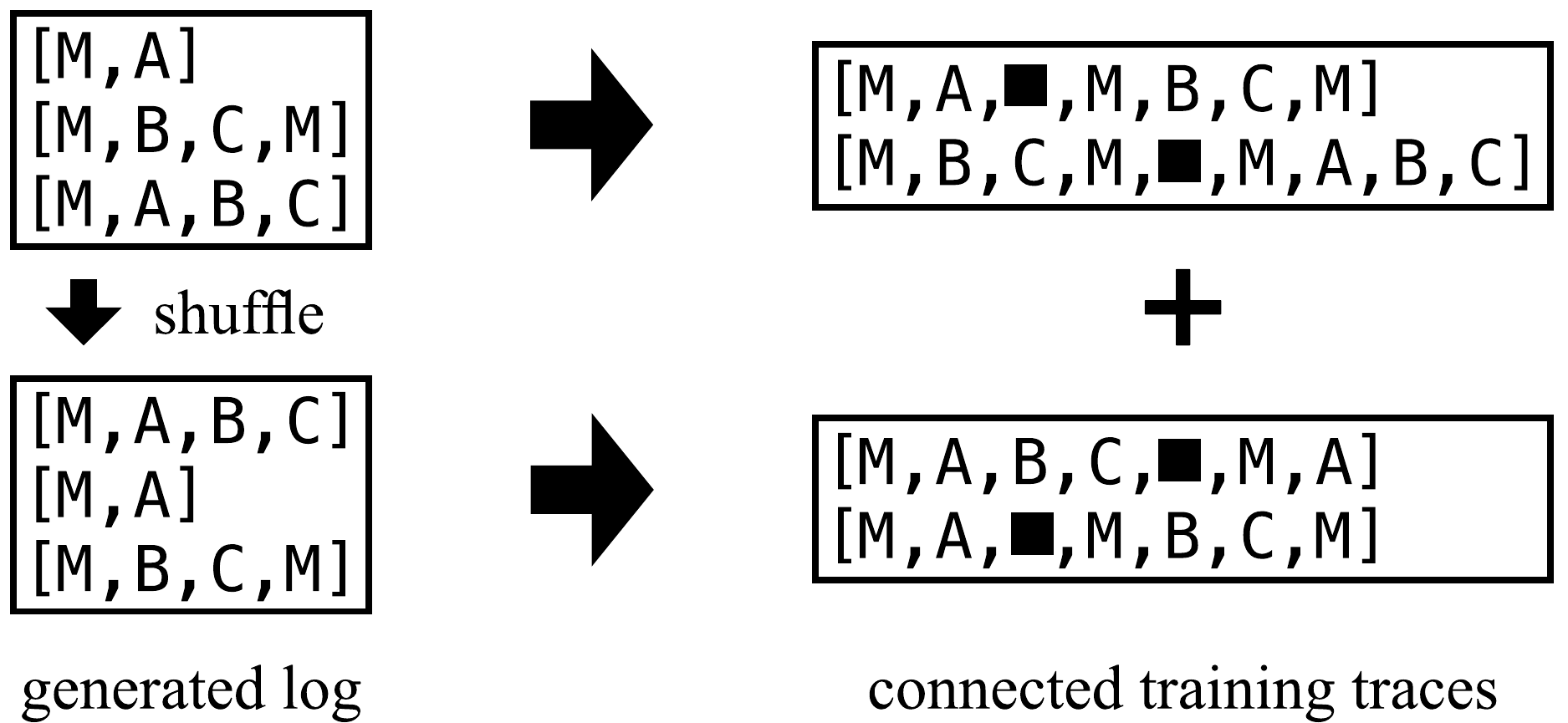}
		\caption{Construction of the training instances. Traces are shuffled and concatenated with a placeholder end activity.}
		\label{fig:training}
	\end{minipage}\hfill
	\begin{minipage}[t]{0.45\textwidth}
		\centering
		\includegraphics[width=\textwidth]{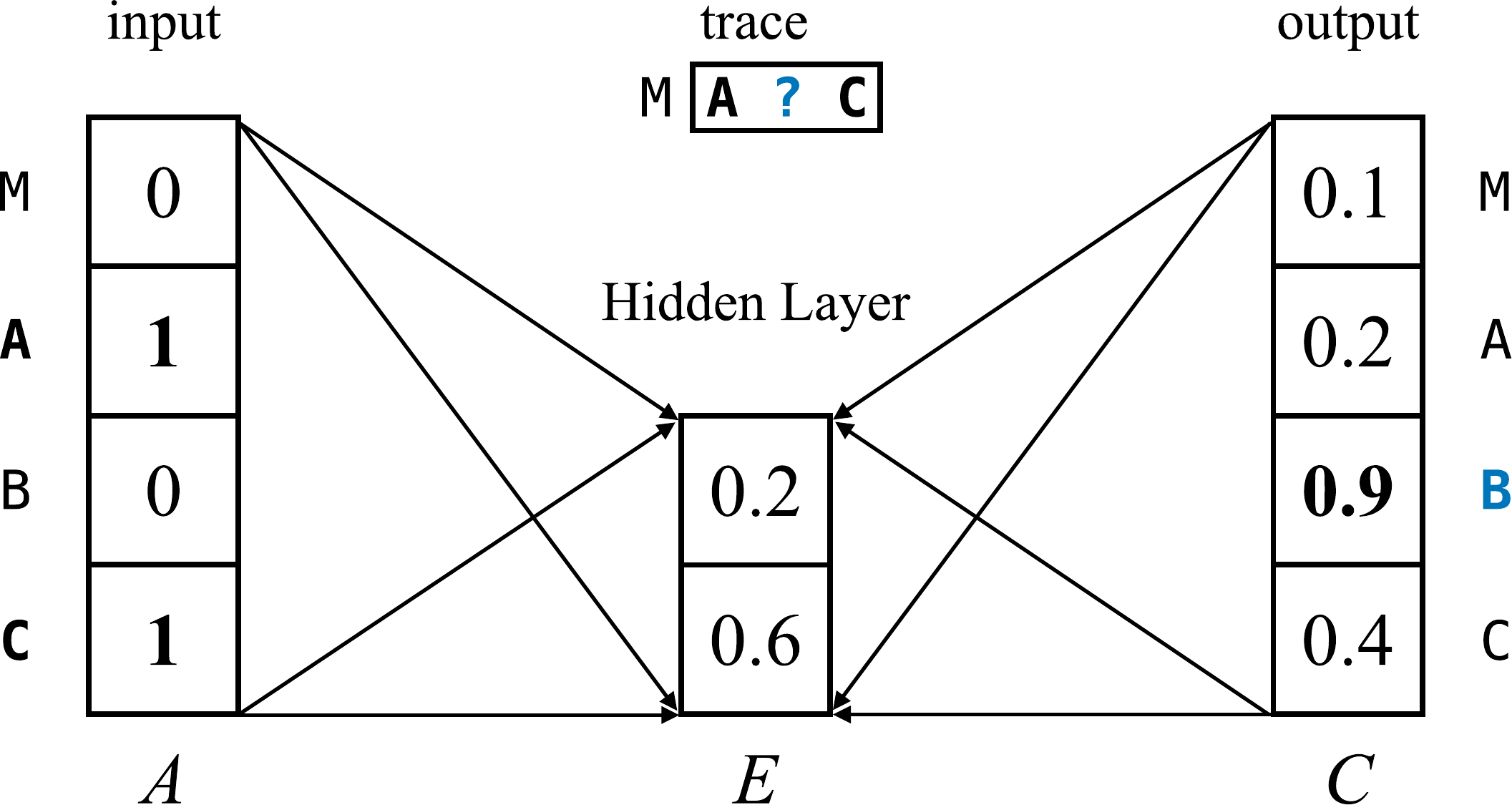}
		\caption{The word2vec neural network. Given the sequence $\langle A, ?, C \rangle$, the network produces a probability distribution over the possible activity labels for $?$.}
		\label{fig:net}
	\end{minipage}
\end{figure}

The training log $L_T$ needs an additional processing step to be used as training set for word2vec. Given two traces $\sigma_1 \in L_T$ and $\sigma_2 \in L_T$, we build a training instance by joining them in a single sequence, concatenating them with a placeholder activity $\blacksquare$. So, for instance, the traces $\sigma_1 = \langle a_1, a_2, a_4, a_5 \rangle \in L_T$ and $\sigma_2 = \langle a_6, a_7, a_8 \rangle \in L_T$ are combined in the training sample $\langle a_1, a_2, a_4, a_5, \blacksquare, a_6, a_7, a_8 \rangle$. This is done repeatedly, shuffling the order of the traces. Figure~\ref{fig:training} shows this processing step on the running example.

The word2vec model~\cite{DBLP:conf/nips/MikolovSCCD13} consists of three layers: an input layer, a single hidden layer, and the output layer. This model has already been successfully employed in process mining to solve the problem of missing events~\cite{DBLP:conf/smc/LakhaniN19}. During training, the network reads the input sequences with a sliding window. The activity occupying the center of the sliding window is called the \emph{center action}, while the surrounding activities are called \emph{context actions}. The proposed method uses the \emph{Continuous Bag-Of-Words} (CBOW) variant of word2vec, where the context actions are introduced as input in the neural network in order to predict the center action. The error measured in the output layer is used for training in order to adjust the weights in the neural network, using the backpropagation algorithm. These forward and backward steps of the training procedure are repeated for all the positions of the sliding window and all the sequences in the training set; when fully trained, the network will output a probability distribution for the center action given the context actions. Figure~\ref{fig:net} shows an example of likelihood estimation for a center action in our running example, with a sliding window of size 3.

\subsection{Segmentation}
\begin{figure}[]
	\centering
	\includegraphics[width=.5\textwidth]{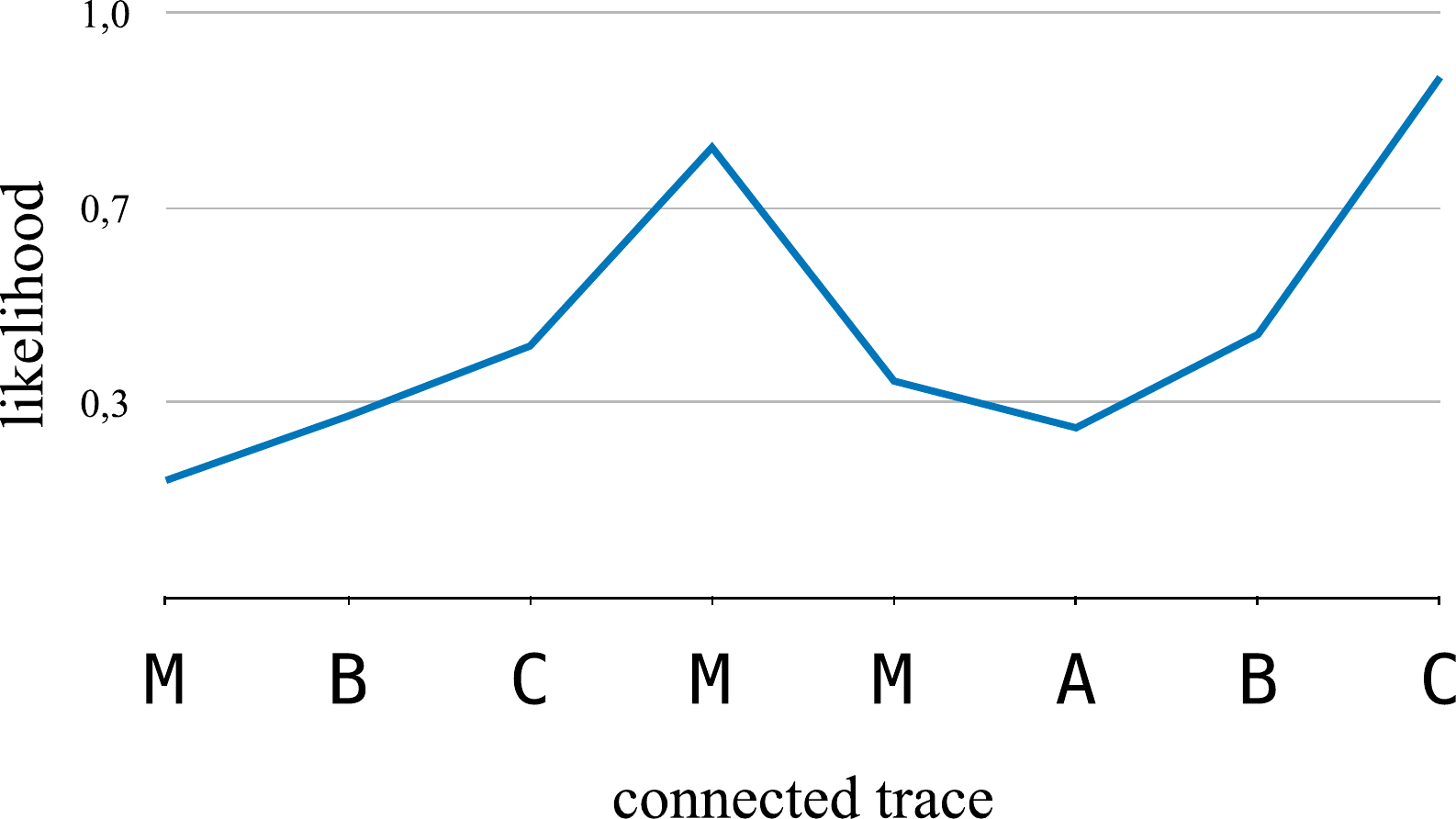}
	\caption{A plot indicating the chances of having a case segment for each position of the user interaction data (second and third trace from the example in Figure~\ref{fig:dfg}).}
	\label{fig:segment}
\end{figure}

Through the word2vec model we trained in Section~\ref{sec:training}, we can now estimate the likelihood of a case boundary $\blacksquare$ at any position of a sequence of user interactions. Figure~\ref{fig:segment} shows these estimates on one user interaction sequence from the running example. Note that this method of computing likelihoods is easy to extend to an ensemble of predictive models: the different predicted values can be then aggregated, e.g., with the mean or the median.

Next, we use these score to determine case boundaries, which will correspond to prominent peaks in the graph. Let $\langle p_1, p_2, \dots, p_n \rangle$ be the sequence of likelihoods of a case boundary obtained on a user interaction sequence. We consider $p_i$ a boundary if it satisfies the following conditions: first, $p_i > b_1 \cdot p_{i-1}$; then, $p_i > b_2 \cdot p_{i+1}$; finally, $p_i > b_3 \cdot \frac{\sum_{j=i-k-1}^{i-1} p_j}{k}$, where $b_1, b_2, b_3 \in [1, \infty)$ and $k \in \mathbb{N}$ are hyperparameters that influence the sensitivity of the segmentation. The first two inequalities use $b_1$ and $b_2$ to ensure that the score is sufficiently higher than the immediate predecessor and successor. The third inequality uses $b_3$ to make sure that the likelihood is also significantly higher than a neighborhood defined by the parameter $k$.

These three conditions allow us to select valid case boundaries within user interaction sequences. Splitting the sequences on such boundaries yields traces of complete process executions, whose events will be assigned a unique case identifier. The set of such traces then constitutes a traditional event log, ready to be analyzed with established process mining techniques.

\section{User Study}\label{sec:qual}
In order to validate the utility of process mining workflows in the area of user behavior analysis, a case study was conducted. Such study also aims at assessing the quality of the segmentation produced by the proposed method in a real-life setting, in an area where the ground truth is not available (i.e., there are no normative well-formed cases). We applied the proposed method to a dataset which contains real user interaction data collected from the mobile applications of a German vehicle sharing company. We then utilized the resulting segmented log to analyze user behavior with an array of process mining techniques. Then, the results were presented to process experts from the company, who utilized such results to identify critical areas of the process and suggest improvements.

\begin{figure}[t]
	\centering
	\includegraphics[width=.65\textwidth]{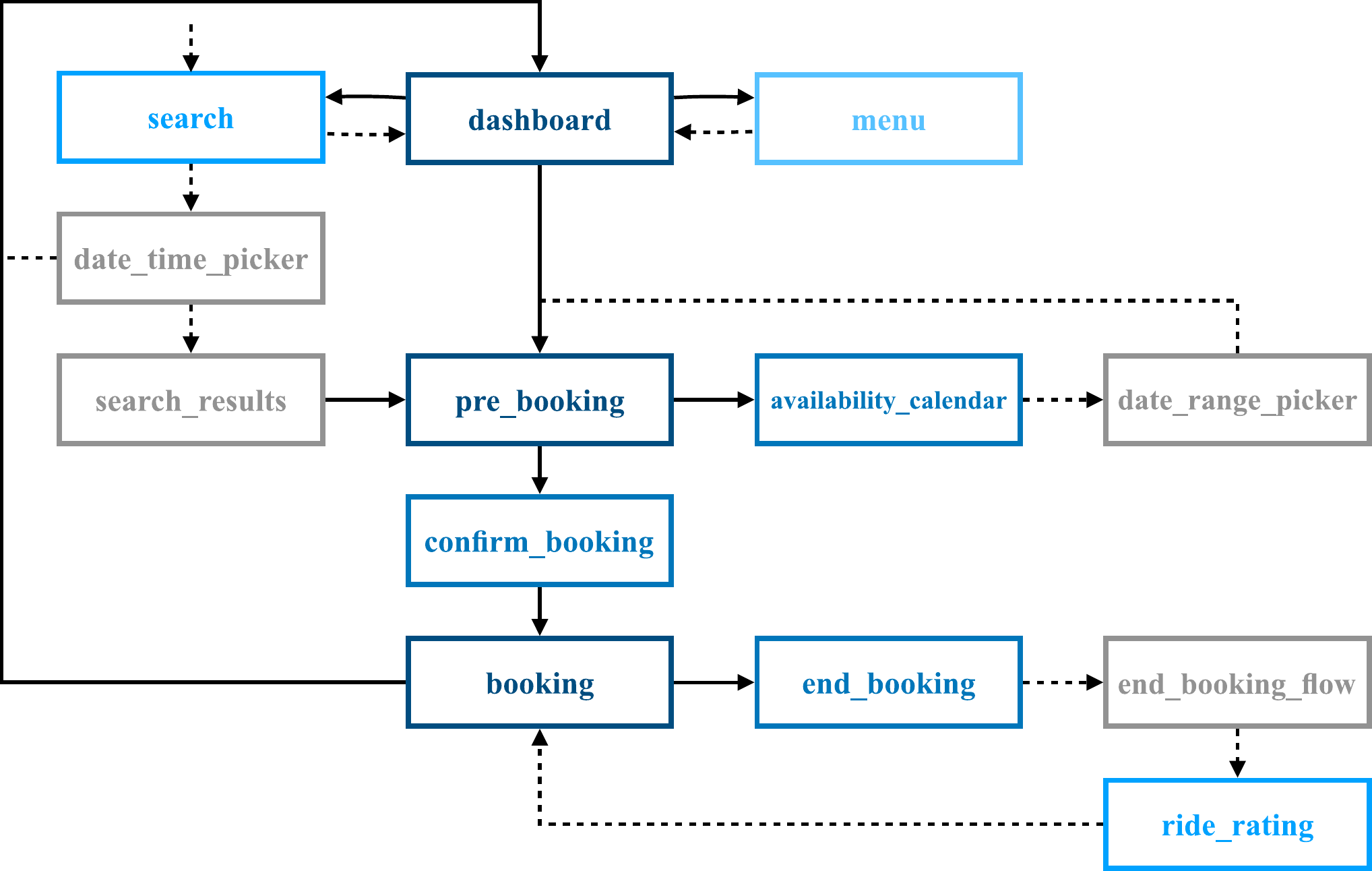}
	\caption{DFG automatically discovered from the log segmented by our method.}
	\label{fig:true}
\end{figure}

In the data, the abstraction for recorded user interactions is the screen (or page) in the app. For each interaction, the system recorded five attributes: \texttt{timestamp}, \texttt{screen}, \texttt{user}, \texttt{team}, and \texttt{os}. The \texttt{timestamp} marks the point in time when the user visited the screen, which is identified by the \texttt{screen} attribute, our activity label. The \texttt{user} attribute identifies who performed the interaction, and the \texttt{team} attribute is an additional field referring to the vehicle provider associated with the interaction. Upon filtering out pre-login screens (not associated with a \texttt{user}), the log consists of about 990,000 events originating from about 12,200 users. A snippet of these click data was shown in Table~\ref{table:data}, in Section~\ref{sec:introduction}.

We applied the segmentation method presented in Section~\ref{sec:method} to this click data. We then analyzed the resulting log with well-known process mining techniques. Lastly, the findings were presented to and discussed with four experts from the company, consisting of one UX expert, two mobile developers and one manager from a technical area. All of the participants are working directly on the application and are therefore highly familiar with it. We will report here the topics of discussion in the form of questions; for reasons of space, we will only document a selection of the most insightful questions.\\

\noindent \textbf{Q1: Draw your own process model of the user interactions.}

\begin{figure}[t]
	\centering
	\includegraphics[width=.6\textwidth]{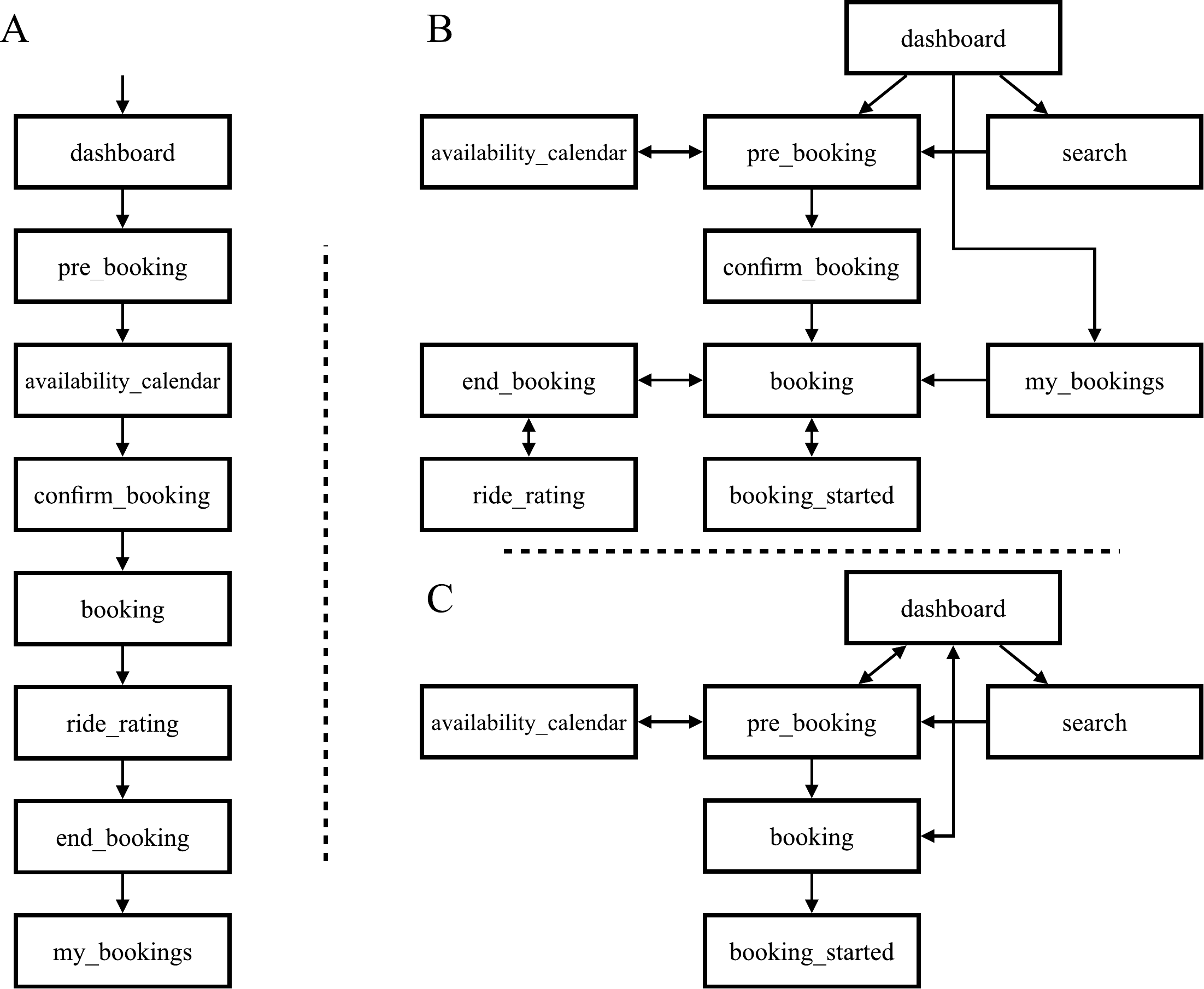}
	\caption{DFGs created by three of the process experts as part of Q1.}
	\label{fig:expert1}
\end{figure}

\begin{figure}[h!]
	\centering
	\includegraphics[width=.6\textwidth]{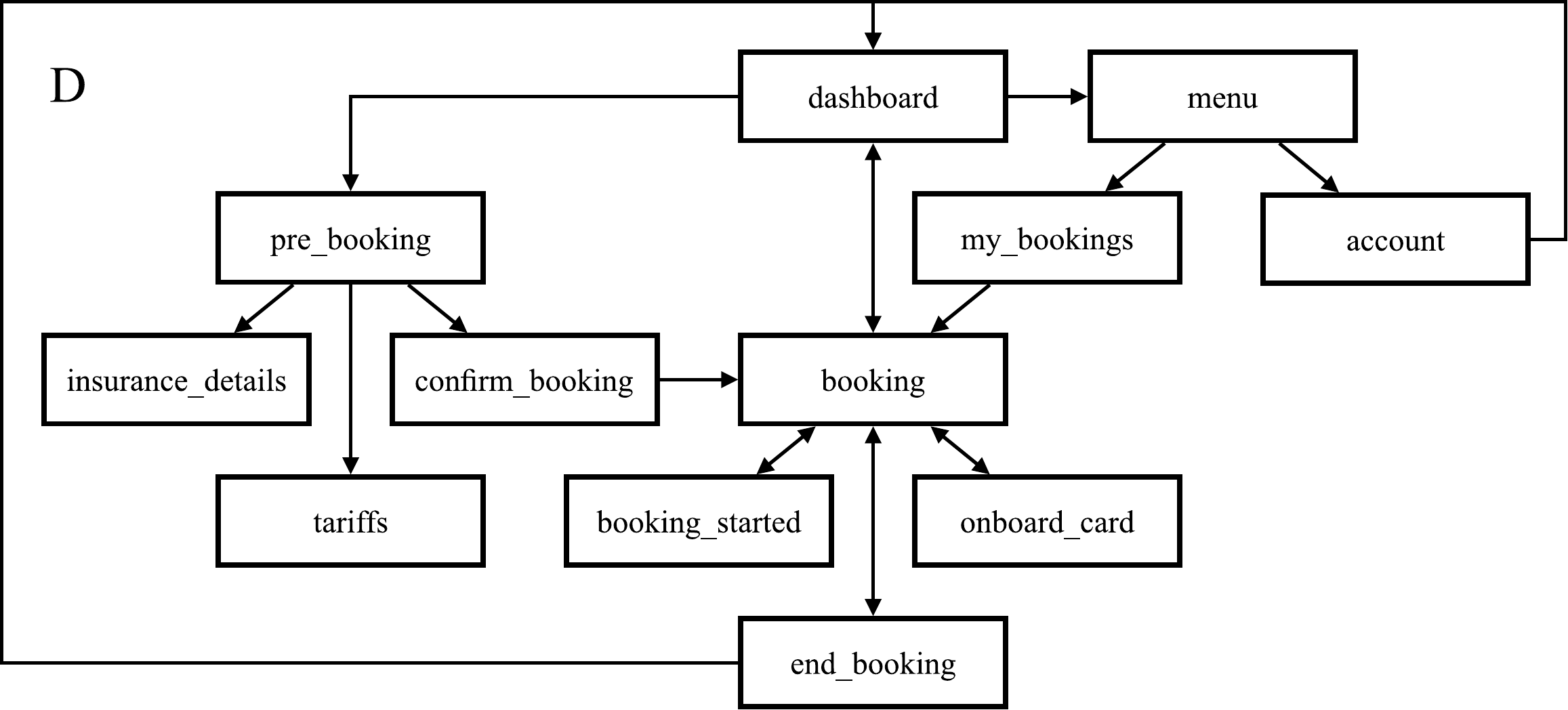}
	\caption{DFG created by one of the process experts as part of Q1.}
	\label{fig:expert2}
\end{figure}

\noindent The participants were asked to draw a \emph{Direcly-Follows Graph} (DFG) describing the most common user interactions with the app. A DFG is a simple process model consisting in a graph where activities A and B are connected by an arc if B is executed immediately after A. The concept of this type of graph was explained to the participants beforehand. The experts were given five minutes in order to create their models. A cleaned up representation of the resulting models can be seen in Figures~\ref{fig:expert1} and~\ref{fig:expert2}.

For comparison, we created a DFG of the segmented log (Figure~\ref{fig:true}). Such model was configured to contain a similar amount of different screens as the expert models. The colors indicate the agreement between the model and the expert models. Darker colors signify that a screen was included in more expert models. The dashed edges between the screens signify edges that were identified by the generated model, but are not present in the participant's models.

The mobile developers (models A and B) tend to describe the interactions in a more precise way that follows the different screens more closely, while the technical manager and UX expert (C and D) provided models that capture the usage of the application in a more abstract way. The fact that the computed model and the expert models are overall very similar to each other suggests that our proposed method is able to create a segmentation that contains cases that are able to accurately describe the real user behavior.\\

\noindent \textbf{Q2: Given this process model that is based on interactions ending on the \texttt{booking} screen, what are your observations?}

\begin{figure}[t]
	\centering
	\includegraphics[width=.5\textwidth]{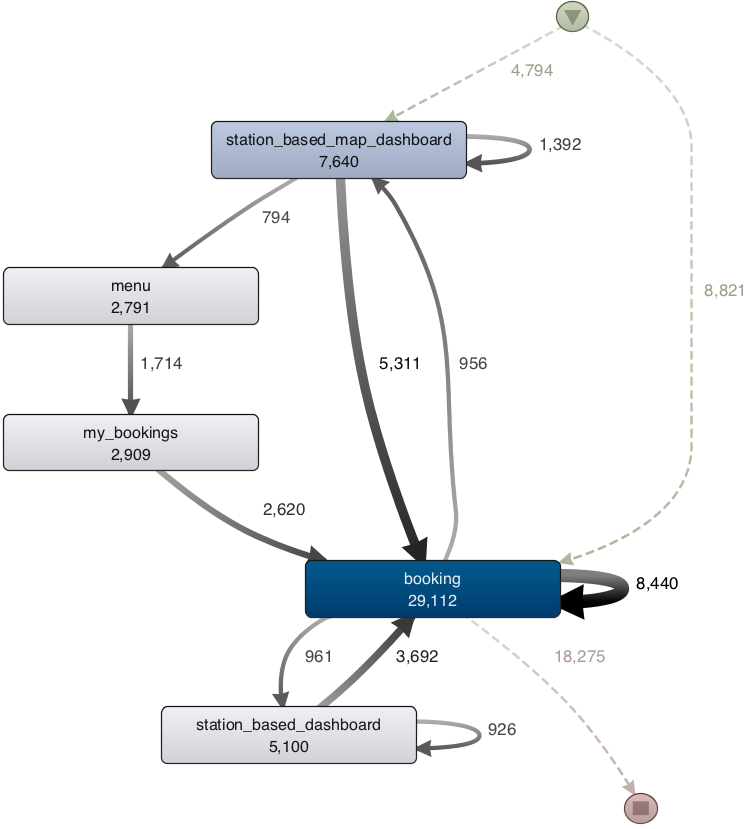}
	\caption{A process model created using Disco, with the \texttt{booking} screen as endpoint of the process.}
	\label{fig:discofreq}
\end{figure}

\noindent Given the process model shown in Figure~\ref{fig:discofreq}, the participants were surprised by the fact that the map-based dashboard type is used significantly more frequently than the basic dashboard is surprising to them. Additionally, two of the experts were surprised by the number of users that are accessing their bookings through the list of all bookings (\texttt{my\_bookings}). This latter observation was also made during the analysis of the segmented log and is the reason that this process model was presented to the experts. In general, a user that has created a booking for a vehicle can access this booking directly from all of the different types of dashboards. The fact that a large fraction of the users take a detour through the menu and booking list in order to reach the booking screen is therefore surprising. This circumstance was actually already identified by one of the mobile developers some time before this evaluation, while they were manually analyzing the raw interaction recordings data. They noticed this behavior because they repeatedly encountered the underlying pattern while working with the data for other unrelated reasons. Using the segmented user interaction log, the behavior was however much more discoverable and supported by concrete data rather than just a vague feeling. Another observation that was not made by the participants is that the path through the booking list is more frequently taken by users that originate from the map-based dashboard rather than the basic dashboard. The UX expert suspected that this may have been the case, because the card that can be used to access a booking from the dashboard is significantly smaller on the map-based dashboard and may therefore be missed more frequently by the users. This is a concrete actionable finding of the analysis that was only made possible by the use of process mining techniques in conjunction with the proposed method.\\

\noindent \textbf{Q3: What is the median time a user takes to book a vehicle?}

\noindent The correct answer to this question is 66 seconds. This was calculated based on the median time of all cases in which a vehicle booking was confirmed. Three participants gave the answers 420 seconds, 120 seconds and 120 seconds. The fourth participants argued that this time may depend on the type of dashboard that the user is using and answered 300 seconds for the basic dashboard and 120 seconds for the map-based dashboard. When asked to settle on only one time, the participant gave an answer of 180 seconds. Overall this means that the experts estimated a median duration for this task of 3 minutes and 30 seconds. This again is a significant overestimation compared to the value that was obtained by analyzing the real user behavior. Again, a mismatch between the perception of the experts and the real behavior of the users was revealed.\\

\begin{figure}[]
	\centering
	\includegraphics[width=\textwidth]{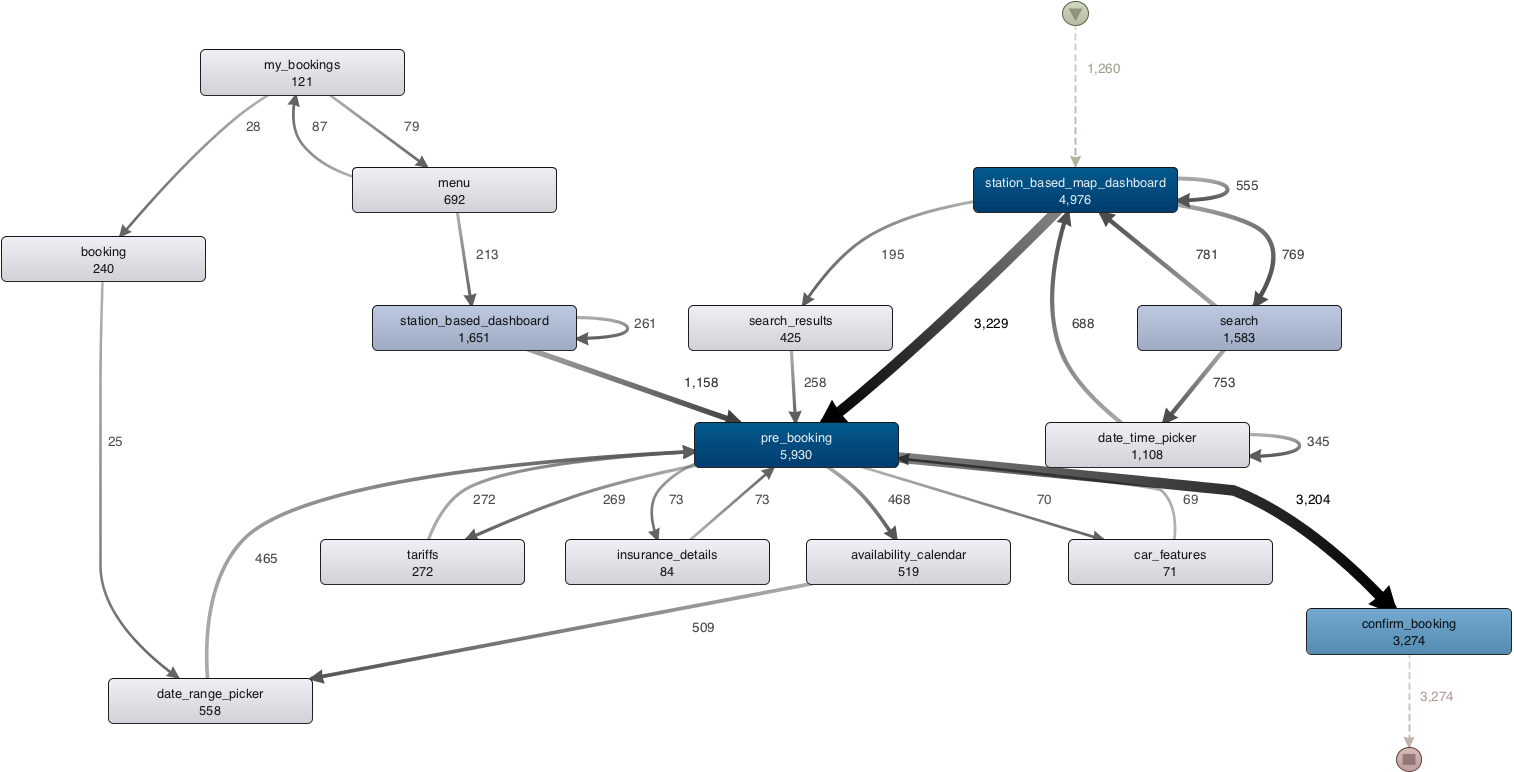}
	\caption{A process model based on cases that begin in any dashboard and end on the \texttt{confirm\_booking} screen.}
	\label{fig:modelconfirm}
\end{figure}

\noindent \textbf{Q4: Given this process model that is based on interactions ending on the \texttt{confirm booking} screen (Figure~\ref{fig:modelconfirm}), what are your observations?}

\noindent Several of the experts observed that the screens that show details about the vehicles and the service, such as \texttt{tariffs}, \texttt{insurance\_details} and \texttt{car\_features}, are seemingly used much less frequently than expected. In only about 2-10\% of cases, the user visits these screens before booking a vehicle. When considering the concrete numbers, the \texttt{availability\_calendar} screen (which is used to choose a timeframe for the booking) and the \texttt{tariffs} screen (which displays pricing information) are used most frequently before a booking confirmation. This suggests that time and pricing information are significantly more important to the users than information about the vehicle or about the included insurance. These findings sparked a detailed discussion between the experts about the possible reasons for the observed behavior. Nonetheless, this shows that models obtained from segmented user interaction logs are an important tool for the analysis of user behavior and that these models provide a valuable foundation for a more detailed analysis by the process experts. Another observation regarding this model was, that a majority of the users seem to choose a vehicle directly from the dashboard cards present on the app rather than using the search functionality. This suggests that the users are more interested in the vehicle itself, rather than looking for any available vehicle at a certain point in time.\\

\noindent \textbf{Q5: Discuss the fact that 2\% of users activate the intermediate lock before ending the booking.}

\noindent The smartphone application offers the functionality to lock certain kinds of vehicles during an active booking. This is for example possible for bicycles, which can be locked by the users during the booking whenever they are leaving the bicycle alone. To do so, the \texttt{intermediate\_lock} and \texttt{intermediate\_action} screens are used. During the analysis, it was found that 2\% of users use this functionality in order to lock the vehicle directly before ending the booking. This is noteworthy, as it is not necessary to manually lock the vehicle before returning it. All vehicles are automatically locked by the system at the end of each booking. One expert argued that this may introduce additional technical difficulties during the vehicle return, because the system will try to lock the vehicle again. These redundant lock operations, discovered analyzing the segmented log, may introduce errors in the return process.\\

\noindent \textbf{Q6: Discuss the fact that only 5\% of users visit \texttt{damages} and \texttt{cleanliness}.}

\noindent The application allows users to report damages to the vehicles and rate their cleanliness, through the homonymous pages. It was possible to observe that only a small percentage of the users seem to follow this routine, which was surprising to the experts. For the vehicle providers it is generally important that the users are reporting problems with the vehicles; optimally, every user should do this for all of their bookings. According to the data, this is however not the case, as only a small percentage of the users are actually using both of the functionalities. The experts, therefore, concluded that a better communication of these functionalities is required.

\section{Conclusion}\label{sec:conclusion}
In this paper, we showed a case and user study on the topic of the problem of event-case correlation. This classic process mining problem was presented here in the specific domain of application of user interaction data.

We examined a case study, the analysis of click data from a mobility sharing smartphone application. To perform log segmentation, we proposed an original technique based on the word2vec neural network architecture, which can obtain case identification for an unlabeled user interaction log on the sole basis of a link graph of the system as normative information. We then presented a user study, where experts of the process were confronted with insights obtained by applying process mining techniques to the log segmented using our method. The interviews with experts confirm that our technique helped to uncover hidden characteristics of the process, including inefficiencies and anomalies unknown to the domain knowledge of the business owners. Importantly, the analyses yielded actionable suggestions for UI/UX improvements. This substantiates both the scientific value of event-log correlation techniques for user interaction data, and the validity of the segmentation method presented in this paper.

Many avenues for future work are possible. The most prominent one is the need to further validate our technique by lifting it from the scope of a user study by means of a quantitative evaluation, to complement the qualitative one showed in this paper. Our segmentation technique has several points of improvement, including the relatively high number of hyperparameters: thus, it would benefit from a heuristic procedure to determine the (starting) value for such hyperparameters. Lastly, it is important to consider additional event data perspectives: one possibility, in this regard, is to add the data perspective to the technique, by encoding additional attributes to train the neural network model.

\bibliographystyle{splncs04}
\bibliography{bibliography}

\begin{thebibliography}{10}
\providecommand{\url}[1]{\texttt{#1}}
\providecommand{\urlprefix}{URL }
\providecommand{\doi}[1]{https://doi.org/#1}

\bibitem{DBLP:journals/sosym/AalstRVDKG10}
van~der Aalst, W.M.P., Rubin, V.A., Verbeek, H.M.W., van Dongen, B.F., Kindler,
  E., G{\"{u}}nther, C.W.: Process mining: a two-step approach to balance
  between underfitting and overfitting. Software and Systems Modeling
  \textbf{9}(1),  87--111 (2010)

\bibitem{DBLP:conf/caise/BayomieAE16}
Bayomie, D., Awad, A., Ezat, E.: Correlating unlabeled events from cyclic
  business processes execution. In: Advanced Information Systems Engineering -
  28th International Conference, CAiSE 2016, June 13-17, 2016. Proceedings.
  Lecture Notes in Computer Science, vol.~9694, pp. 274--289. Springer (2016)

\bibitem{DBLP:conf/er/BayomieCRM19}
Bayomie, D., Ciccio, C.D., Rosa, M.L., Mendling, J.: A probabilistic approach
  to event-case correlation for process mining. In: Conceptual Modeling - 38th
  International Conference, {ER} 2019, November 4-7, 2019, Proceedings. Lecture
  Notes in Computer Science, vol. 11788, pp. 136--152. Springer (2019)

\bibitem{DBLP:journals/dke/BurattinKNW19}
Burattin, A., Kaiser, M., Neurauter, M., Weber, B.: Learning process modeling
  phases from modeling interactions and eye tracking data. Data \& Knowledge
  Engineering  \textbf{121},  1--17 (2019)

\bibitem{DBLP:conf/bpm/FerreiraG09}
Ferreira, D.R., Gillblad, D.: Discovering process models from unlabelled event
  logs. In: Business Process Management, 7th International Conference, {BPM}
  2009, September 8-10, 2009. Proceedings. Lecture Notes in Computer Science,
  vol.~5701, pp. 143--158. Springer (2009)

\bibitem{DBLP:conf/icpm/JanssenMKZ20}
Janssen, D., Mannhardt, F., Koschmider, A., van Zelst, S.J.: Process model
  discovery from sensor event data. In: Process Mining Workshops - {ICPM} 2020
  International Workshops, October 5-8, 2020, Revised Selected Papers. Lecture
  Notes in Business Information Processing, vol.~406, pp. 69--81. Springer
  (2020)

\bibitem{DBLP:conf/IEEEscc/JlailatyGB17}
Jlailaty, D., Grigori, D., Belhajjame, K.: Business process instances discovery
  from email logs. In: 2017 {IEEE} International Conference on Services
  Computing, {SCC} 2017, June 25-30, 2017. pp. 19--26. {IEEE} Computer Society
  (2017)

\bibitem{DBLP:conf/smc/LakhaniN19}
Lakhani, K., Narayan, A.: A neural word embedding approach to system trace
  reconstruction. In: 2019 {IEEE} International Conference on Systems, Man and
  Cybernetics, {SMC}, October 6-9, 2019. pp. 285--291. {IEEE} (2019)

\bibitem{DBLP:conf/icpm/LenoADRMP20}
Leno, V., Augusto, A., Dumas, M., Rosa, M.L., Maggi, F.M., Polyvyanyy, A.:
  Identifying candidate routines for robotic process automation from
  unsegmented {UI} logs. In: 2nd International Conference on Process Mining,
  {ICPM} 2020, October 4-9, 2020. pp. 153--160. {IEEE} (2020)

\bibitem{DBLP:conf/gi/LinnZW18}
Linn, C., Zimmermann, P., Werth, D.: Desktop activity mining - {A} new level of
  detail in mining business processes. In: 48. Jahrestagung der Gesellschaft
  f{\"{u}}r Informatik, Architekturen, Prozesse, Sicherheit und Nachhaltigkeit,
  {INFORMATIK} 2018 - Workshops, September 26-27, 2018. {LNI}, vol. {P-285},
  pp. 245--258. {GI} (2018)

\bibitem{DBLP:conf/ACMdis/MarrellaC18}
Marrella, A., Catarci, T.: Measuring the learnability of interactive systems
  using a {Petri Net} based approach. In: Proceedings of the 2018 on Designing
  Interactive Systems Conference, {DIS}, June 09-13, 2018. pp. 1309--1319.
  {ACM} (2018)

\bibitem{DBLP:conf/nips/MikolovSCCD13}
Mikolov, T., Sutskever, I., Chen, K., Corrado, G.S., Dean, J.: Distributed
  representations of words and phrases and their compositionality. In: Advances
  in Neural Information Processing Systems 26: 27th Annual Conference on Neural
  Information Processing Systems. Proceedings of a meeting held December 5-8,
  2013 (2013)

\bibitem{DBLP:journals/kais/MurillasRA20}
de~Murillas, E.G.L., Reijers, H.A., van~der Aalst, W.M.P.: Case notion
  discovery and recommendation: automated event log building on databases.
  Knowledge and Information Systems  \textbf{62}(7),  2539--2575 (2020)

\bibitem{DBLP:conf/icpm/0001BUA21}
Pegoraro, M., Bakullari, B., Uysal, M.S., van~der Aalst, W.M.P.: Probability
  estimation of uncertain process trace realizations. In: Munoz{-}Gama, J., Lu,
  X. (eds.) Process Mining Workshops - {ICPM} 2021 International Workshops,
  October 31 - November 4, 2021, Revised Selected Papers. Lecture Notes in
  Business Information Processing, vol.~433, pp. 21--33. Springer (2021)

\bibitem{DBLP:journals/is/PegoraroUA21}
Pegoraro, M., Uysal, M.S., van~der Aalst, W.M.P.: Conformance checking over
  uncertain event data. Information Systems  \textbf{102},  101810 (2021)

\bibitem{DBLP:conf/apn/PegoraroUA21}
Pegoraro, M., Uysal, M.S., van~der Aalst, W.M.P.: {PROVED:} {A} tool for graph
  representation and analysis of uncertain event data. In: Application and
  Theory of Petri Nets and Concurrency - 42nd International Conference, {PETRI}
  {NETS} 2021, June 23-25, 2021, Proceedings. Lecture Notes in Computer
  Science, vol. 12734, pp. 476--486. Springer (2021)

\bibitem{DBLP:journals/ijcis/PourmirzaDG17}
Pourmirza, S., Dijkman, R.M., Grefen, P.: Correlation miner: Mining business
  process models and event correlations without case identifiers. International
  Journal of Cooperative Information Systems  \textbf{26}(2),
  1742002:1--1742002:32 (2017)

\bibitem{DBLP:conf/caise/RamirezR0V19}
Ramirez, A.J., Reijers, H.A., Barba, I., Valle, C.D.: A method to improve the
  early stages of the robotic process automation lifecycle. In: Advanced
  Information Systems Engineering - 31st International Conference, CAiSE, June
  3-7, 2019, Proceedings. Lecture Notes in Computer Science, vol. 11483.
  Springer (2019)

\end{thebibliography}

\end{document}